## THE NEVER ENDING SEARCH FOR HIGH TEMPERATURE

## SUPERCONDUCTIVITY

THEODORE H GEBALLE

Departments of Applied Physics and Materials Science

Stanford University, Stanford CA 94305

August 15, 2006

## I. HISTORICAL INTRODUCTION

It is a pleasure to contribute to this book and honor Professor Vitaly Ginzburg who has been a major contributor to the field of superconductivity for over half a century. He has been steadfast in his belief that higher temperature superconductivity is possible and has proposed model structures in which it might be found. I have shared his long-standing interest and have searched experimentally for novel superconductors and for higher limits of $T_c$s. In the best of circumstances theoretical and experimental approaches should complement each other.

For the first four decades after Kammerlingh Onnes discovered superconductivity [1] research was carried out in the few laboratories that had access to liquid helium; research was mainly confined to a few elements and alloys [2]. At the time when the Ginzburg-Landau theory [3] was formulated in 1950 superconductors were laboratory curiosities and were limited in number. Of course important work had been done and much of the basic thermodynamic and electromagnetic properties of what are now known as type-I superconductors had been established [4], but little was understood about the underlying microscopics [5]. There was no basis for predicting the occurrence of superconductivity, and there was little apparent connection with the normal metallic state, even though the properties of simple metals were quite well described by the quantum theory of electrons in metals. Outstanding physicists including Einstein, Bohr, Bloch, and Heisenberg tried, but were unable to find a satisfactory microscopic theory of superconductivity until BCS in 1956 [6].

The rare and unpredictable occurrence of superconductivity and the lack of theory led Enrico Fermi at the University of Chicago in ~1950 to encourage two young colleagues, John Hulm and Bernd Matthias, to undertake a broad search for new superconductors [I joined a few years later.] Fermi may have been influenced by the strange discoveries of Hans Meissner, who in the 1929 had found the barely metallic compound CuS [7, 8] to be superconducting in contrast with metallic Cu that was not. Soon thereafter Meissner found a number of superconducting intermetallic borides and carbides [9, 10]. During this period most of the few laboratories equipped with the necessary cryogenic infrastructure for investigating superconductivity were engaged in studying the macroscopic properties of pure superconducting elements. The motivation for searching for new superconductors and the materials science expertise for doing so that may have originated in Hans Meissner's laboratory in the 1920's and 1930's was reborn with renewed vigor first at the University of Chicago and soon was transferred to the Bell Labs and Westinghouse Research Laboratories by Matthias and Hulm respectively. They discovered many new superconducting alloys and intermetallic compounds [11]. The Periodic Table was found to be a valuable guide for predicting the occurrence and magnitude of $T_c$ using what came to be known as Matthias' rules. The most useful "rule" is simply an empirically determined dependence of $T_c$ upon the algebraic average number of the valence electrons per atom in the 3d, 4d, and 5d series of

transition metal elements and compounds - i.e. to a crude average of the electron density

[12]. To a considerable extent as a consequence of Matthias' "rules", superconductivity

changed from being a rare phenomenon to being common. References to the many new superconductors that were found during this period are given in the compilation of B. W.

Roberts [13].

As the database increased, refinements were incorporated; superconductivity was

found to be favored in specific structures [14], the most intriguing being the A15

structure (also known as beta tungsten and $Cr_3Si$ [15]). Compounds with this structure

had the highest known $T_c$s up until 1986 and also possessed low temperature martensitic

transitions and marked phonon softening. These have been associated with the unusual

arrangement of orthogonal and non-intersecting chains of closely spaced atoms of

transition metals and a Fermi surface with weakly one dimensional features [16]. The

discovery of the superconductivity of $V_3Si$ [17] soon led to the $T_c = 18.1K$ of $Nb_3Sn$ [18]

and opened a new age of superconductivity. Its dawn was not fully appreciated until

Kunzler, Wernick and coworkers discovered that $Nb_3Sn$ could carry huge currents in high

fields [19]. This high-field high-current capability forced the abandonment of the

Mendelssohn sponge theory [see discussion in ref 2] that had up until that time had

discouraged experiments with so-called "hard" superconductors because it (incorrectly)

attributed high critical fields to the presence of very thin filaments that were incapable of carrying large currents. Abrikosov's already existing but not widely appreciated theory of quantized flux and type-II superconductivity was soon "discovered" and provided the correct explanation [20, 21]. The science and technology that emerged from the discovery of $Nb_3Sn$ is a striking example of how the discovery of a new superconductor can create a rich new field of research. Searching for new superconductors is a high risk endeavor, but with some luck and a prepared mind (Pasteur) new doors can be opened and hopefully there will be more occurrences in the future [see Appendix A]. Today due to marked advances in thin film growth and characterization techniques there are opportunities for reaching far beyond the limitations of bulk phase equilibria. It is possible to make systematic investigations of artificial structures with monolayer control of chemical composition, with novel interfaces, with strain-induced epitaxy, and with various temperature and pressure cycles. There have also been advances in methods for the controlled growth of single crystals, as well, that have been facilitated by the development of computer controlled growth techniques [22].

There is nothing to compare with the impact made by the discovery of High Temperature Superconductivity [23] in the layered perovskite related cuprates. While there have been many models proposed for the microscopic superconducting (pairing) interactions there is still no consensus for the mechanism. Prof Ginzburg has ranked "high-temperature and room-temperature superconductivity" second in his list of 30 "especially important and interesting problem as of 2001" [24].

I believe, contrary to the most practitioners who believe that the pairing interactions are confined solely to the $CuO_2$ layers, that until interactions throughout the unit cell are understood we will not have a satisfactory theory that can account for $T_c$s > ~100K. In the past there have been many of reports of signals attributed to superconductivity at high temperatures, even well above room temperature, particularly since Bednorz and Mueller's discovery [25]. Most, but perhaps not all, are probably spurious and can be attributed to poor experimental practice. They have not been duplicated elsewhere; the signals are usually not reproducible and tend to disappear with time. It is conceivable, however, in some cases that micro-amounts of metastable superconducting regions exist. These might percolate through regions that are unstable or metastable and give rise to the observed ephemeral macroscopic electric or magnetic superconducting signals. Various kinds of stacking faults, nanoscale compositional segregation or other defects may be responsible for a few of the reports of $T_c$ above 200K in the layered cuprate systems. In the appendix we cite a few cases as examples of systems that may be worth rechecking under more controlled conditions.

## II. SUPERCONDUCTING MECHANISMS

### Phonons

As early as 1922 Kammerlingh Onnes and Tuyn looked for a difference in "the vanishing point " of [the resistivity of] Pb and uranium Pb, i.e. $^{206}$Pb [26]. Unfortunately the gas thermometer they used was reproducible to 0.01K, just about the same size as the signal they should have seen. It was not until 1950 that the isotope effect was established [27]. The electron-phonon mechanism was incorporated in the famous BCS theory six years later. In 1959 after Gorkov derived the Ginzburg-Landau equations from BCS it became possible to test the theory and find that it could explain and make quantitative predictions of a wide variety of phenomena. A direct experimental "proof" for the phonon mechanism was provided by the conductance curves of superconducting-insulating-normal metal (SIN) tunnel junctions that were analyzed using Eliashberg's equations [28] that yielded a (weighted) phonon density of states that closely matched the phonon density of states determined by inelastic neutron scattering. In BCS theory the electron pairing can be mediated by any bosonic degree of freedom that exists within the system. In all the investigations made before 1986 that I know of, only phonons were identified. Thallium doped PbTe that is discussed below is most likely an exception due to an exciton mechanism.

**Non-phonon pairing**

$T_c$ is limited by the phonon energy scale (the prexponential factor in the BCS expression) and can only be approached within a factor of ~10 in the strong coupling limit [29]. This limitation stimulated attempts to synthesize structures that incorporate exciton-mediated pairing. Predictions of much higher $T_c s$ were based upon the much higher exciton energy scales ~1eV vs. phonons energies ~0.02eV. However while the bosonic energy enters explicitly in the preexponential factor of the strong coupling version of BCS, it also enters in to the coulomb interaction that is in the exponent because of retardation. [30]. $T_C = \omega \cdot \exp(-\frac{1+\lambda}{\lambda - \mu^*})$, Where $\lambda$ is the electron-boson coupling constant and $\mu^* = \frac{\mu}{1 + Log(E_f / \omega)}$ is the retarded coulomb repulsion. However, there can be little or no retardation in homogeneous systems when the boson and band energies are of the same order. As a consequence the full (unscreened) coulomb repulsion is felt. This will reduce $T_c$, in fact if it is larger than the attractive interaction it will completely suppress the superconductivity. To overcome this problem models were proposed in which the electrons and excitons are spatially separated. Little proposed a one-dimensional conductor with polarizable side chains [31], Ginzburg proposed a two-dimensional system across a metal-dielectric interface where the exciton-induced pairing in a thin metal film is due to the penetration of polarization waves of and from the underlying dielectric [32]. Allender, Bray, and Bardeen [33] proposed a somewhat different two-dimensional model where the metal electrons tunnel into a polarizable semiconductor. These models have been investigated experimentally and no evidence for excitonic superconductivity has been found. It has been shown theoretically that under some assumptions they are unphysical [34]. However Ginzburg and Kirzhnits [35] found that there are some restricted conditions where excitonic pairing across two-dimensional interfaces is possible.

In the next section we present evidence for excitonic superconductivity in a much different system than those discussed above, namely the semiconductor PbTe in which ~ 1% of the Pb is replaced by Tl. The pairing interaction is believed to be localized within the Tl ionic volume and the system can thus in some sense be considered to be a 0-dimensiona superconductor.

**Negative-U paring[a]**

It is well known from chemistry that Tl and Bi are valence skipping elements, that is, in solids they occur as $6S^0$ and $6S^2$ ions but not with the intermediate $6S^1$ configuration. Compounds in which they might be expected to have paramagnetic $6S^1$ configuration are found to be diamagnetic due to disproportionation, for example TlS where — $2Tl^{2+} \rightarrow Tl^{1+} + Tl^{3+}$. This reaction is of course energetically impossible in the vapor phase although correlation does reduce the energy cost of having two electrons in the large 6S orbit. Disproportionation occurs in the solid mainly because of the relaxation and polarizability of the near neighbor sulfur ions. Anderson [36] introduced the concept of a negative-U center to explain why there were no paramagnetic states (as evidenced by the

---

a Parts of the discussion in this and the following sections have been taken from recently submitted papers [82] .

lack of EPR signals) in chalcogenide glasses. In his model, in the low frequency limit, the Fermi level is pinned by the high density of two electron states within the mobility gap of the one electron states. He noted that the lattice relaxation around a localized ion with an unpaired electron could overcome the repulsive coulomb energy of an added second electron, making the effective coulomb energy $U$ be negative. The superconductivity, which was discovered by Chernik and Lykov [37] upon substituting a small concentration of Tl on the Pb site in semiconducting PbTe, is believed to be due to pairing caused by negative-U center Tl ions. Extensive research that has been reviewed by Kaidanov and Ravich [38] and by Nemov and Ravich [39] supports this idea. Hall data suggest that Tl initially substitutes for $Pb^{+2}$ in the lattice of the ionic semiconductor as a shallow $Tl^{+1}$ acceptor located in the valence band. Upon further doping up to concentrations slightly above 0.3% the samples become superconducting. PbTe can be doped by many other acceptors over the same concentration range but only Tl induces superconductivity. Moyzhes and Suprun [40] have proposed a model in which the charge fluctuations that occur on the Tl ion are screened by the high frequency dielectric constant. Thus the pairing is mainly electronic in origin. A recent investigation by Matsushita *et al.* [41] finds low temperature resistance minima occur in the same concentration range as the superconductivity. The accompanying theory by Derzo and Schmalien [42] shows that their data can be fit with the charge-Kondo model. The key assumption of the charge Kondo model is that the scattering center has degenerate internal degrees of freedom that in the present case would be the two charge states of Tl. The degeneracy is confirmed by Hall measurements. For concentrations below ~0.3% each Tl acceptor contributes one hole to the valence band thus lowering the Fermi level but neither a resistance minimum nor superconducting transition are observed. Above ~0.3% and the Hall coefficient becomes much less dependent upon Tl concentration because the [41] Tl self-compensates, ( disproportionates) forming the +1 and +3 states that pin the Fermi level and both resistance minima and superconductivity are observed. Evidently the localized Tl- ions exchange charge in units of 2e with the extended valence band states. Below $T_c$ pairing becomes coherent presumably by proximity coupling of the Tl ions through the valence band states. Analysis of the superconducting properties shows that Pb(Te Tl) is a weak coupled superconductor [43]. The fact that the observed $T_c s$ are two orders of magnitude higher than the $T_c s$ of comparable chemically similar semiconductor-superconductors [44] shows that the pairing energy scale is much higher. This is indirect evidence for negative-U pairing that is quite distinct from the one- and two-dimensional excitonic models proposed by Little, Ginzburg and Bardeen mentioned above. Because the localized 6S orbital of the Tl ion is the smallest length scale in the system, Pb(TeTl) is a quasi 0-dimensional negative-U center superconductor.

The above discussion raises the question-- can negative-U ions be pairing centers in the high $T_c$ cuprates? It is known that the highest $T_c$ cuprates all contain charge reservoir layers that are oxides of negative-U ions. This suggests that the superconductivity found in the highest $T_c$ cuprates is due to enhancement by negative-U centers as discussed below. It further suggests that a good strategy for searching for new high temperature superconducting systems is to find structures with negative-U centers where the chemical potential can be controlled so as to make the ionic configurations degenerate.

## III. PAIRING AND $T_c$ IN CUPRATE SUPERCONDUCTORS

Perovskite and perovskite-related structures have complex unit cells containing 3 or more sites and are found to have a rich assortment of ordered ground states [45]. Among them are the high-$T_c$ layered cuprates all of which contain 2-dimensional layers of $CuO_2$. This common feature has influenced most theories to start with the not unreasonable assumption that the superconductivity arises from the interactions within the 2-dimensional $CuO_2$ layers. With this assumption numerous pairing models have been proposed. They account for the increase in $T_c$ in any given family of cuprates when n, the number of $CuO_2$ layers within unit cell, increases from 1 to 3 in terms of quantum tunneling [46]. Decreases in $T_c$ with further increases in $n$ will be discussed below.

My aim in this chapter is modest. It is to show, by comparing $T_c s$ in *different* cuprate families, that the pairing in the $CuO_2$ layers must be supplemented by interactions elsewhere in the unit cell. This conclusion is reached simply by considering the significant variations in $T_c$ that are found in structures that have the same sequence of $CuO_2$ layers within the unit cell but have different intervening layers (the columns in Table 1).

**The ionic model**

It is convenient to start from the insulating side using a simple ionic model because it provides an intuitive approach to a complex problem. The ionic model has long been used as a basis for understanding insulating oxides and their magnetic properties. It is a limit of strong correlation and thus has some credibility as an approximation for underdoped cuprate superconductors. In the Born approximation the large attractive Madelung energy is balanced by the repulsive overlap energy. It is useful, following Moyzhes *et al*. [40] to modify the Born equation and to use the high frequency dielectric constant to account for the electronic polarizability of the electron clouds of the near neighbor ions, and to use the low frequency dielectric constant to account for the motion of more distant ions. This modification has lead to a statistically significant classification of a large number of oxides, and to make useful predictions as to stability vs. instability, and metallic vs. insulating character to within about 1 eV [47]. The modified Born equation is a crude but non-trivial representation of the local density (LDA) approximation [48].

**The Cu ion**

Before proceeding to discuss the cuprates it is worthwhile to recall some chemistry that makes the Cu ion unique. In the vapor phase $Cu^{+2}$ has the highest 3rd ionization potential of the transition metals. This large energy is retained in the condensed state as is evident from the electrode potentials of ions in aqueous solution [49]. Electrode potentials provide rough estimates of the relative ionic energies in crystalline oxides because in both the aqueous and crystalline environments the cations are coordinated by oxygen ions. The standard electrode potential for charge transfer $Cu^{+3} + e^- = Cu^{+2}$, $E(0) = +2.4eV$ is very high. It follows that in cuprates the doped holes will reside mainly on oxygen sites (in contrast to other transition metal oxides where the cations are oxidized upon hole doping). On the other hand, the standard electrode potential for the reaction $Cu^{+2} + e^- = Cu^{+1}$ is quite low, $E(0) = -0.15eV$, showing that $Cu^{+2}$ can easily coexist with $Cu^{+1}$. Consequently in the condensed state $Cu^{+1}$ and $Cu^{+2}$ are close in energy. Of course in the crystalline cuprates, the band, the exchange, the correlation and the crystal field energy must be included. But the ionic energies are the largest .We can assume without further calculation that $Cu^{+3}$ (d8 configuration) does not play a significant role in the dynamics of the cuprates, and that the cuprates are "charge transfer insulators". The distinction between charge transfer insulators and Mott insulators was first noted by Zaanen, Sawatsky and Allen [50].

**The CuO$_2$ layers**

In the undoped parent compounds the CuO$_2$ layers are insulating antiferromagnets containing Cu$^{+2}$ ions. The magnetic moment of the Cu ions is substantially less than the 1 Bohr magneton expected for divalent Cu ion showing that there is a redistribution of the formal charge due to overlap of the half-filled Cu ($x^2$-$y^2$) d-levels with the O$^{-2}$ that results in a localized narrow lower Hubbard band [51]. Doping of holes in the CuO$_2$ layers can be accomplished in 3 ways: by chemical substitution of cations with lower valences (Sr for La), or by cation reduction (Tl$^{+3}$ to Tl$^{+1}$), or by the addition of oxygen ions. As already discussed the added holes reside mainly on the oxygen p-levels. These are more extended than the Cu levels, and the antiferromagnetism is rapidly destroyed. When the hole concentration in La$_2$CuO$_4$ (214) exceeds ~0.05 holes per Cu the samples become superconducting [52]. $T_c$ follows a dome shaped curve as a function of doping. The maximum $T_c$ is reached at the optimum concentration p = 0.16 holes/Cu for (LaSr)$_2$CuO$_4$ [53]. It has been commonly assumed that the optimum concentration is strictly a property of the CuO$_2$ layer and thus is a "universal" property of all the cuprate superconductors. However, $T_c$ depends upon coupling between the layers and as we argue below the coupling between CuO$_2$ layers is not universal so there is no reason to expect a universal optimum concentration. Thus the results of Karppinen that $T_c$ of Bi-2212 occurs for $p$ = 0.12 should not be surprising [54].

**Negative-U paring in charge reservoir layers**

Important deductions can be made by comparing optimum $T_c$s of cuprates that have the same number $n$ of CuO$_2$ layers per unit cell, but have different layers separating them (columns in Table 1). The intervening layers that contain the cations Tl, Bi and Hg ions are known as charge reservoir layers because they exchange charges with the $CuO_2$ layers. Tl, Bi, negative-U ions exchange charge in units of $2e$. Hg is a two-center negative-U ion that also exchanges charge in units of $2e$ [55]. This leads to the hypothesis that the negative-U ions become a coherent part of the pair field and enhance $T_c$. The empirical evidence is straightforward.[56]. In 214 family of cuprates (based upon $La_2CuO_4$) $T_c$ reaches a $T_{c(max)} \sim 40K$ when the $CuO_2$ layer is optimally doped by substitution of Sr for La, or up to 45K when the doping is by interstitial oxygen ions staged in more remote layers, or up to 51 K in strained epitaxial films. What must be explained is the mechanism by which the $T_c$ of these "optimally" doped 214 cuprates is increased to >90K simply by inserting charge reservoir layers containing oxides of Tl, Hg or Bi. The separation of the $CuO_2$ layers which is 6.6 Å in the 214 compound is increased by 3 (5) Å by the insertion of one (two) TlO layers which by itself alone should only decrease $T_c$.

The hypothesis that the negative-U centers can enhance $T_c$ gains credibility by a theory [57], exact in the weak coupling limit, that shows negative-U centers can be resonant pair tunneling centers when incorporated in the barrier of a Josephson junction. In any real junction (to our knowledge none have been investigated) the negative-U centers within the barrier will most likely interact with each other to form clusters in order to maximize overlap with each electrode [57]. There is obviously a strong analogy between clusters of negative-U centers between the electrodes in a Josephson junction and the negative-U centers in the charge reservoir layers between the $CuO_2$ layers.

Table 1: Variation in $T_c$

| CuO$_2$/$_e$ | n=1 | | n=2 | | n3 | |
|---|---|---|---|---|---|---|
| | $T_c$(K) | Separations (Å) | $T_c$(K) | Separations (Å) | $T_c$(K) | Separations (Å) |
| LSCO-214 | 40 | 6.6 | - | - | - | - |
| Hg-12(n-1)n | 98 | 9.5 | 127 | 9.5 | 134 | 9.5 |
| Tl-12(n-1)n | - | | 103 | - | 133 | - |
| Tl-22(n-1)n | 95 | 11.5 | 118 | 11.5 | 125 | 11.5 |
| Bi-22(n-1)n | 38 | - | 96 | - | 120 | - |
| Y123 (6 GPa) | - | - | 95 | 7.9 | - | - |
| Y124 (6 GPa) | - | - | 105 | 9.8 | - | - |

## IV. EVIDENCE FOR NEGATIVE-U PAIRING CENTERS

### The Tl cuprates

T. Suzuki et al. [58] have observed by XPS spectroscopy that the Tl $4f_{7/2}$ core level valence in Tl-2223 lies roughly midway between the core levels of the +3 and +1 reference standard oxides Tl$_2$O$_3$.and Tl$_2$O. Since the nondisproportionated +2 ionic configuration should be at a much higher energy, the measured value most likely represents a superposition of the Tl+1 and +3 configurations that is averaged over a time scale shorter than the measurement time and that exchanges charge with the CuO$_2$ layers in units of 2e. Terada and coworkers have found further such evidence in studies of Tl-1223 also from core level spectroscopy. When the $T_c$ of Tl-1223 is increased from 100K to above 130K upon annealing in vacuum the rather broad peak of the Tl $4f_{7/2}$ core level shifts from being centered at Tl$^{+3}$ (determined by the standard Tl$_2$O$_3$) to being centered half-way between Tl$^{+3}$ and Tl$^{+1}$ (the Tl$_2$O standard) [59]. The correlation between the >30K $T_c$ rise and the shift of the Tl ion valence follows from the doping of holes (in units of 2e) into the CuO$_2$ layers for the same reason given above. We thank Norio Terada for pointing out an alternative possibility. If the as-prepared Tl-1223, with $T_c$ = 100K, initially was overdoped (a condition that I beilieve is unlikely because it is only the double TlO layered compounds that accommodate interstitial oxygen) , then the removal of oxygen from the TlO layer by annealing could account for both the increased $T_c$ and the Tl valence reduction. In either case the enhanced superconductivity and an increase in negative-U centers are correlated, but further experiments to clarify this interesting result are called for.

**The Hg Cuprates**

The mercury cuprates are interesting for several reasons beyond having the highest known $T_c > 160K$ found in the Hg-1223 compound under pressure. The homologous series $HgBa_2Ca_{n-1}Cu_nO_{2n+2+\delta}$ has been synthesized [60] all the way from $n = 1$ to $n=7$ with $T_c$ for the optimally doped samples rising from 97K at n = 1 to maximum at $n = 3$ and then decreasing for $n>3$.

The increase from $n = 1$ to $n = 3$ is common to all the cuprates (Table 1) as we have already mentioned follows from the coupling over the short distance between the $n$ layers within the unit cell for example by quantum tunneling [61]. Subsequent decreases with $n > 3$ can be understood from NMR investigations that show that the $CuO_2$ layers are not uniformly doped [62]. Mukuda [63] find in optimally doped $n = 5$ samples that the inner layers are antiferromagnetic with $T_N = 60K$ and with ~0.35 Bohr magnetrons per Cu [64], while the outer layers are superconducting with $T_c = 108K$. The robust persistence of the superconductivity indicated by negative curvature of the dependence of $T_c$ upon $n$ for $n > 3$ would be surprising if the superconducting interactions were confined only to the single outer $CuO_2$ layers. In a somewhat comparable model system [65] that contains two $CuO_2$ layers sandwiched between antiferromagnetic undoped $La_2CuO_4$ layers, $T_c$ is only 30K.

The much higher~ 80K found in Hg cuprates is more easily understood if the HgO layers
are also contributing to the pairing.

$T_c$ rises universally with pressure for the $n = 1$ to $n = 3$ Hg cuprates with the same
unusually large coefficient, $dT_c/dP \sim 2.0$ K/GPa, for all three. The pressure coefficient is
constant as a function of doping from low level to optimum for the Hg-1201 and even
remains positive in the overdoped region of Hg1212 [66]. As already noted all three have
the same pressure dependence when compared at optimal doping) [ 67 ]. The
independence of $dT_c/dP$ upon concentration is difficult to be explained by the models that
assume charge transfer to be responsible for the record high $T_c$s reached at high pressure.
Such models would predict a steadily decreasing pressure coefficient in going from under
doped to optimally doped where it should approach zero unless there are some quite
special compensating changes in the band structure with pressure. However, the behavior
follows from the negative-U model because pressure should increase the overlap of the
pairing centers in the HgO layers with the $CuO_2$ layers. Such overlap follows from
possibly orbital overlap of the Cu d-$z^2$ orbitals with the apical oxygen [68].

The HgO-BaO layers are highly disordered [69]. There are a large number of oxygen
vacancies in the HgO layers. XAFS measurements of the Hg-Hg distances are of such
poor quality that they cannot be modeled [70]. Consequently the negative-U centers in
the HgO layers are probably more complex than the idealized two center ion discussed in
[56] and in the pair tunneling model [57] it is reasonable to assume that clusters of
interacting negative-U Hg centers form in the HgO-BaO layers.

**The Bi Cuprates**

Karppinen *et al.* [54] find by cation doping that the $T_{c(max)}$ for Bi-2212 is reached when the hole concentration in the $CuO_2$ layers is $\sim 0.12$ holes/Cu (Fig. 1) as determined by chemical titration and independently by analysis of the Cu $L_{2,3}$- absorption edge. Only half of the doped holes appear in the adjacent $CuO_2$ layers, an almost equal amount are found in the BiO layers.

In the Emery-Kivelson model [71] $T_c$ is limited as a function of doping in the $CuO_2$ layers by the lesser of either the pairing energy or phase stability. The optimum concentration is consequently determined by the intersection of curves representing pair amplitude (that decreases as a function of hole doping) and phase stiffness that increases with superfluid density (hole doping). In that model there is no a priori reason why the optimum concentration should be fixed at the 0.16 value that has been determined for the 214 family although 0.16 is frequently assumed to be a universal property of the $CuO_2$ layer. Any interaction that stiffens the phase fluctuations, as the negative-U ion pairing will do, will drive the optimum concentration to lower values. We attribute Karppinen's result [54] to just that effect as sketched in the phase diagram Fig. 2. $T_{c(max)}$ occurs at nearly the same 1/8 concentration that is known to suppress superconductivity in the 214 cuprates due to competition from charge ordering [71]. Evidently the pairing in the BiO layers strengthens the superconductivity so that it competes successfully with the two dimensional charge ordered state.

## V. THE CHAIN LAYER CUPRATES

There is much evidence showing that pairing also occurs in the single and double CuO chain layers of the cuprates.. The chain layers consist of either single CuO chains (123 cuprates) or double ("zigzag") CuO chains (124, or 248 cuprates), or an alternating sequence of single and double layers (247 cuprates.) In all three structures the quasi-one dimensional chains are separated from the blocks of $n=2$ $CuO_2$-(Y,Pr)-$CuO_2$ layers by layers of BaO. An important difference is that the oxygen ion concentration in the single chains of 123 cuprates is variable and permits oxygen doping of the $CuO_2$ layers, in contrast to the 124 cuprates that have a fixed stoichiometry. Doping of the $CuO_2$ layers of course is possible by cation substitution on the Y site.

**Evidence from NQR**

The most direct evidence for pairing in the chain layers is provided by the NMR investigations by Sasaki *et al.* [72] that revealed that the superconductivity discovered by M. Matsukawa *et al.* [73] in Pr247 originates in the double chains layers. As can be seen in Fig. 3, the 247 structure is composed of alternating Pr123 and Pr124 sub-units. Neither of the sub-units by themselves has been found to be superconducting; there is a good understanding for why [74]. As initially prepared by sintering, Pr247 also is not superconducting. However, it undergoes a transition to zero resistance at ~10K when annealed in vacuum at 400ºC. The NQR resonances associated with the four Cu sites in the Pr247 structure are well resolved [75] allowing them to be followed separately. The $CuO_2$ layers order antiferromagnetically around 280K as they do in Pr123 and Pr248. The square root temperature dependence of the relaxation data of the Cu nuclei in the double chains above $T_c$ is evidence of the one dimensional non-Fermi liquid expected for Tomonaga-Luttinger liquid [76] (Fig. 4). The fact that the sharp increase slope occurs at nearly the same temperature at which the resistance goes to zero can hardly be coincidental - it is strong evidence for superconductivity that originates in the chains. Analysis [72] indicates that only 20% of the chain copper nuclei contribute which is not surprising considering the fragile nature of one dimensional superconductivity. Perhaps much higher $T_c$s would be reached if better samples could be prepared. The path through which the double chains become coherent is probably through the antiferromagnetic $CuO_2$ layers via the apical oxygen ions [68] as discussed below in connection with Zn substitutions in the Pr-cuprates.

While ~10K is not "high temperature" in comparison with other cuprates, it is very high when compared with other strongly 1D systems such as $(SN)_x$, $T_c = 0.3K$ [77]. A linear diamagnetic quasiparticle that is stabilized by the coulomb energy gained when doped hole binds to a charge transfer exciton has been proposed to account for the transport in the double chains [78]. Such a linear configuration has previously been introduced by Shklovskii and Efros as the ground state in compensated semiconductors [79]. How the annealing that removes oxygen from the single chain layers results in the superconductivity in the double chain layers is an interesting question that deserves further investigation. Comparable annealing experiments of the Pr124 double chain cuprates have shown no such effects.

**Evidence from anisotropy**

The CuO chains running in the b-crystal direction must be directly or indirectly responsible for the considerable planar anisotropies observed in transport and in the superconducting penetration depths. In optimally doped single chain Y123 the planar anisotropy is ~1.8 for d.c. and optical transport in the normal state, and probably not coincidentally it is the same for the superfluid density in the superconducting state as determined from penetration depth measurements (assuming any effective mass change is small). In Y124 the anisotropy is even greater, ~3 in both the normal and superconducting state even though the crystalline planar anisotropy is less in the 248 than in the 123. Models that attribute the superfluid density in the chains to a proximity interaction fail to predict the wide temperature range over which the anisotropy of superfluid density is temperature independent [80, 81]. The most plausible explanation is that there is intrinsic pairing in the chain layers that is of the same origin that we have just seen occurs in the Pr247 double chains. The fact that the same planar anisotropy is found in both the normal and superconducting properties density can be understood simply if the quasiparticles in the normal state form the pairs in the superconducting state, and other effects are not important. If the anisotropy were induced in the $CuO_2$ layers by the chain layers (strain for example) then it should be greater in single chain cuprates because, as noted above, they are more orthorhombic.

**Other evidence**

There is also considerable evidence from the dependence of $T_c$ upon pressure and strain, and from cation substitution. These results that have been discussed elsewhere [82] can be understood if the chain-chain coherence is established not by direct overlap (over the 4Å distance perpendicular to the chain axis) but rather indirectly by coupling through the $CuO_2$ layers. This indirect coupling is suggested by experiments where a few percent of Zn is doped into $CuO_2$ layers. Upon doping, $T_c$ rapidly decreases with almost the same dependence upon concentration in both the Y123 and Y124 cuprates. This alone would be most simply understood if all the superconductivity originated in the $CuO_2$ layers, but such an interpretation is not viable in the light of the following experiments with non-superconducting Pr-124. Here the $CuO_2$ layers are semiconducting and the resistance is understood in terms of parallel conduction paths in metallic chains and in the semiconducting layers; the planar transport anisotropy is 1000 at 4K [83]. With a few percent Zn substitution in the $CuO_2$ layers [84] the a-axis becomes insulating, the b-axis conduction develops the power law dependence characteristic of a one dimensional Luttinger-Tomonaga conductor [76]. ARPES further shows that the Fermi level disappears giving rather direct evidence the chains have become decoupled in the a-direction as a result of the Zn doping [85]. Evidently the localization length in the $CuO_2$ layers becomes shorter than 4 angstroms needed to couple the chains in the *a*-direction. As expected from this model in Y248 the strain coefficient of $T_c$ in the direction perpendicular to the chains, $dT_c/da$, is very large and positive. Comparable strain experiments in the 123 cannot be interpreted because of the complications associated with oxygen diffusion and reordering.

## VI. SUPERCONDUCTIVITY ORIGINATING IN THE $CuO_2$ LAYERS

There are more than enough good theories available to explain the superconductivity in the $CuO_2$ layers. My purpose here is to show that the ionic models used above have value in discussing interactions within the $CuO_2$ layers as well as in the chain and charge reservoir layers. Experimental values of the charge transfer gap are $\leq$ 2eV for $La_2CuO_4$ in the antiferromagnetic state at low temperatures [86]. Gaps in the same energy range are found in $HgBa_2CuO_4$ [87] and likely in all the high $T_c$ cuprates. There is also considerable subgap structure. The lowest peak at $\sim$ 0.4eV is in reasonable agreement with the ionic estimate of the lowest energy charge transfer exciton, taken to be the gap energy minus the screened interaction between the bound charges, giving an energy $E_{ex} = E_g - q^2/\varepsilon r$. Here, $q$ is the absolute value of the charges, $r$ is their separation, and $\varepsilon$ is the dielectric response. Putting $E_g$ = 2eV and $r$ = 2Å and making the reasonable assumption that for the short distance, $\varepsilon = \varepsilon_\infty$ = 5, gives $E_{ex} \sim$ 0.5eV. Some of the subgap structure may also be due to multi-magnon/phonon processes [88]. Upon doping, the bands broaden and the gap edge is lowered. Resonant inelastic X-ray scattering (RIXS) data show that with doping the gap is filled with an electron-hole continuum [89, -- We would like to acknowledge a helpful private communication from Y. Ando.]. In our model this occurs when the polarization cloud of the lowest lying charge transfer exciton (Fig. 5a) binds with the doped hole (Fig. 5b) to form a linear exciton-hole (eh) quasi particle (Fig. 5c). This is the same linear configuration introduced by Shklovskii and Efros [90] as the ground state in compensated semiconductors. The eh particle is a linear charge-one spin-zero quasi particle with an energy estimated in the ionic model as $E = E_{ex} - [q^2/\varepsilon r - q^2/2\varepsilon r] = +0.5eV - 0.72eV = -0.22eV$. The well-known Zhang Rice (ZR) singlet is an alternate configuration that places the doped hole in a symmetrical molecular orbital of the oxygen ions surrounding a given Cu ion [91] and gains considerable exchange energy [92]. On the other hand the eh-singlet is slightly stabilized by coulomb energy and more importantly is a favorable configuration for being a mobile quasiparticle in the double chain layers of the 248 compounds, and for stripe formation in the $CuO_2$ layers [93].

In the limit $t_{pp} = 0$ [Fig. 5e] the electron dynamics are purely one-dimensional and the eh particle would live in either a (10) or (01) configurations. Cluster calculations, however, suggest that $t_{pp}/t_{pd}$ is in the range of 0.3 [94], raising some question about the validity of quasi one-dimensional eh transport. However, the dressed (extended) version of the eh-particle (Fig. 5d), for which $t_{pp}/t_{pd}$ is reduced by a large factor, is favored at low temperatures because of a significant gain in zero-point energy stabilization [95]. At higher temperatures, however, entropy will favor the tri-ion eh-particle. The bent configuration (Fig 5e) has a higher energy due to the larger coulomb interaction $V_{pp}$, between the oxygen ions (Fig 5e).

## VII THE PHASE DIAGRAM

The negative-U addition to the phase diagram in Fig. 2 is qualitatively no different than the generally accepted phase diagrams [96] except that we have allowed for enhancement (doubling) of $T_c$ by negative-U charge reservoirs layers. In the underdoped region below some not-well-defined-temperature $T^*$, well above $T_c$, anomalies (usually called $T^*$ anomalies) are observed in various phenomena such as Knight-shift, spin-lattice relaxation [97], transport and in a reduction of the effective magnetic moments. These are crossover phenomena that in our model are due to the formation of the eh-particles that coexist with the (paramagnetic) doped holes. As the temperature decreases further, but still well above $T_c$. the concentration of eh particles increases to the extent that superconducting fluctuations as observed by Ong and coworkers [98] can be seen. The quasi-one dimensionality of the eh-particles thus leads to fluctuating stripes and charge-spin separation [99]. In this model there is no necessity to postulate separate regions of (01) and (10) domains because of the d-wave symmetry that insures opposite phase relation for stripes in the (01) and (10) directions at the Cu crossing points. There would be no corresponding increase in kinetic energy because of the nodes in the wave functions at the crossing points. However, neutron data do not favor this possibility [100].

We have argued that $T_c$ is enhanced because the negative-U centers increase the superfluid density. A direct way of doing this is simply to double the number of Cu ions by filling the vacant sites in the $CuO_2$ layer and forming $Cu_2O_2$ (i.e., CuO) with double the number of Cu-O bonds. [101]. A thermodynamically stable form of CuO occurs. It is the naturally occurring monoclinic insulator known as tenorite. Real space images of tenorite show evidence of spin-charge separation, and also anisotropic transport is consistent with stripe formation [102].

Finally, as in all models the 3d superconducting transition occurs when the temperature is lowered and the interlayer coupling energy overcomes thermal fluctuations. While d-wave symmetry is favored in the $CuO_2$ layers, a small s-wave component must exist in the chain layer cuprates as a consequence of orthorhombicity, and is also likely to be present in all the cuprates because of disorder. Once a small s-component exists there is no symmetry restriction that prevents coupling to the negative-U ions or negative-U ion clusters.

Acknowledgements I thank Boris Moyzhes for valuable discussions regarding the negative U model in Pb(TeTl), the role of negative U centers as pairing centers, and the eh particle. I would also like to acknowledge valuable discussions with Steven Kivelson and Gertjan Koster covering many of the ideas presented here. I have also benefited from interactions with Mac Beasley ,Ivan Bozovic, Ian Fisher, Aharon Kapitulnik, Z. X. Shen, Doug Scalapino, Dave Blank, Frank DiSalvo, Seb Doniach, Hiroshi Eisaaki, Susuma Sasaki, Norio Terada, Hideki Yamamoto, Yoshi Ando, and Jan Zaanan. I am indebted to Bernd Matthias for instructing me in the art of finding new superconductors and to John Hulm as well, and to Conyers Herring and Phil Anderson for very many helpful insights over the past 5 decades. I thank Yuan Li for a careful reading of this manuscript. It is a pleasure to acknowledge the sustained support of the AFOSR during these years.

## Appendix A--   WHY CONTINUE THE SEARCH?

The purpose of this appendix is to show that searching for new superconductors, in addition the goal of discovering of room temperature superconductivity, leads to novel materials, interesting physics, and the possibility of uncovering new superconducting mechanisms.

nexpected superconductivity is often easy to detect using simple and convenient electrical or magnetic probes. In favorable cases it is possible to discover tiny amounts of a previously unknown compound, or an unanticipated superconducting phase, which is buried in an inhomogeneous sample. Until recently, however, trace quantities had been difficult or even impossible to detect when the superconductivity occurred in isolated islands embedded in an insulating or metallic majority phase. Today a variety of sensitive scanning probes have the potential to do so-- these include tunneling tips, SQUIDS, micro-Hall bridges, near-field microwave probes, magnetic force cantilevers, micropotentiometer probes and perhaps others. Once the presence of the minority-superconducting phase is established it is a "proof of principle" that should challenge materials scientists to isolate the pure phase (as was the case in the early days of the cuprate era).

Ease of detection, however, has been and still is, a two-edged sword. In fortunate situations trace amounts of the superconducting phase can precipitate in grain boundaries and enclose macroscopic regions in connected paths so that shielding currents are easily detected (a positive-positive result). However, signals from non-superconducting inhomogeneous samples can easily mimic superconducting signals, due to a redistribution of current flow caused, for example, by differing temperature-dependent resistivity of the different phases (a positive-negative result). Unfortunately these latter situations seem to be the explanation for most (but perhaps not all) of the reports of near-room- and even higher $T_c$s that have been made sporadically over the years, and have occurred much more frequently after the discovery of high $T_c$. The investigator is most often unable to isolate the suspected superconducting phase or to provide samples to others, and the results have not been reproduced in other laboratories. But there remains a possibility that in some cases the signals are real, and that it comes from an unstable isolated minority phase that may have been produced by an uncontrolled synthesis variable. Possible superconductivity of CuCl under pressure is a good example [103].

**Advanced deposition synthesis**

Today it is possible to synthesize new materials in a previously unattainable vast landscape that has been opened by the remarkable advances made in the techniques of film deposition and characterization. Fine rate-control, and *in situ* and *ex situ* diagnostics make it possible to extend phase boundaries well beyond the thermodynamic limits, to produce new phases made from elements that normally are incompatible, to obtain metastable and strained phases by epitaxial growth on a wide variety of substrates, to produce sharp concentration gradients and charge transfer at interfaces, and to produce arrays in which doping and other synthesis variables can be systematically investigated.

**Interplay between theory and experiment**.

The challenges presented by the discovery of high temperature superconductivity have in the past have extended the forefront of theoretical and experimental research. It is thus reasonable to expect that future discoveries will do likewise. The interplay between theory and experiment is well illustrated by the golden age which opened after the microscopic BCS theory was discovered [104] and the connection between the BCS pair wave function and Ginzburg-Landau phenomenology was established [105]. Seemingly strange results were either explained or became inputs for reaching a deeper understanding of the phenomenon. In some cases strange results turned out to be due to unsuspected material science rather than physics.

It is worthwhile to recall a few well known examples [b].

**Phonon effects**

●The old observation that many poor conductors make good superconductors was clarified [phonons which mediate the superconducting pairing in the superconducting state cause scattering in the normal state.] [c]

●The propensity for $T_c$ to be increasing as a function of composition in a given phase at the limit of its stability. [At the edge of stability mode softening can increase the electron phonon coupling.]

●The variable mass-dependence of $T_c$ [isotope effect] found in transition metals. [time retardation is mass dependent and renormalizes the coulomb pseudopotential.] [106].

**Chemical and Magnetic effects**

●Nb produces higher $T_c$s than other elements, either alone, or as a constituent in the A15 compounds, in the cubic rock salt compounds, and in other intermetallics. That such an observation can be at least rational follows from strong-coupling theory where the pairing interaction is localized in space (and retarded in time [106]).

●The insensitivity of $T_c$ to some impurities and the extreme sensitivity to others. After the initial Leiden discovery of superconductivity in which care was taken to use highly purified mercury it must have been surprising to find that ordinary solder was also

b Other examples are given in White and Geballe (R.M. White and T. H. Geballe;, *Long Range Order in Solids* ,Academic Press (1979)

c $MgB_2$ is a special case of a good superconductor being a good conductor because there is little mixing between the high mobility electrons with the rest of the Fermi surface

"supraconducting" [d] and subsequently that superconductivity is the common ground state of non-magnetic metals [107], even amorphous ones [108]. [Elastic scattering is time invariant and simply renormalizes the basis states by averaging over the Fermi surface [109].]

●On the other hand magnetic impurities cause a rapid decrease in $T_c$ [110]. [spin-spin scattering breaks the time reversal symmetry of the paired electrons in s-wave superconductors [111]. More recent work has shown that the opposite is true for superconducting pairs with p-wave symmetry [112].]

**Materials science**

●There were also cases which apparently contradicted BCS, but further work showed those "failures" were the result of unsuspected materials science . The apparent lack of superconductivity in elemental Mo was due a few hundred ppm of Fe that formed magnetic pair-breaking states [113]. On the other hand dilute concentrations of Fe caused abnormally high $T_c$s in Ti. At first this was considered to due to a new kind of magnetic pairing interaction. Subsequent studies showed that Fe was preferentially highly concentrated in a minority (bcc) phase of Ti that had segregated in connected grain boundaries [114] and the behavior was exactly as expected from the Matthias Rule [115].

**Predictability**

Today there is no theory for the cuprates that has the ability to make quantitative predictions of superconducting properties and in that sense we are still in the equivalent of

---

d The early Leiden communications referred to the phenomena as supra-conductivity, meaning "beyond conductivity" which is a more accurate description than the one in use the pre-BCS era. One method of making progress now, as it was then, is to find new superconductors, particularly in novel conductors with highly correlated systems. Successful searches are usually guided upon an intuitive belief that certain parameters can be brought under experimental control and will play an essential role. The prize example of this type of search is of course Bednorz and Mueller's discovery of cuprate superconductivity [116].

Future discoveries are also likely to come from the "bench top" experiments (small science) as was true in the past. The previously unknown compound $Nb_3Sn$ was made simply by heating the elements in a quartz tube following the earlier discovery of superconductivity in the obscure compound $V_3Si$, Its unanticipated high-field and high critical current capabilities led to new concepts in science that enabled new kinds of electrical power and magnetic technologies such as MRI. The discoveries of superconductivity in heavy-fermion conductors, organic charge-transfer salts, and alkali metal intercalated. C60 (buckyballs) are also examples where scientific fields have been opened. The discoveries of unexpected superconductivity in $MgB_2$, hydrated $Na_xCoO_2$, and $SrRu_2O_4$, are recent examples

**Future**

We have already mentioned the powerful new methods for synthesizing and investigating thin films that have recently become available and are constantly being improved. These open opportunities for investigating vast new landscapes of materials. The odds are good that the more unexpected discoveries will be made and they have not been mentioned in this article. It is possible they will be found in materials that have pairing interactions understandable with present day concepts such as doped Mott insulators, light elements with strong chemical bonds and high Debye temperatures that can be doped using field effect or interface phenomena, by negative-U ions or other kinds of excitons. .

There also may be clues buried among the numerous reports of superconducting-like signals coming from inhomogeneous samples that may or may not be real for reasons discussed at the beginning of this appendix. From these I have arbitrarily selected three classes.

●The first is the report of Osipov *et al* [117] who deposited thin films of elemental Cu on single-crystal insulating CuO (tennorite) substrates. Upon imposing a series of sharp ~10 microsecond current pulses of a few amps , superconducting-like (diamagnetic and electrical) signals were observed at temperatures as high as 400K, High critical currents were estimated using a somewhat uncontrolled estimate of thickness. In similarly uncontrolled experiment [118] silver films deposited on single crystal substrates of $Sr_2CuO_4$ ($T_c$ = 1.4K were recrystallized by laser ablation. Subsequent resistance measurements were reported to give superconducting-like signals above 200K, It is of course necessary to make more controlled experiments and to eliminate the possibility of spurious signals in order to establish its presence of superconductivity that presumably would exist as a quenched metastable surface layer or interface.

●The second class is the subgoup of cuprate superconductors in which weak magnetic and/or resistive signals that are possibly due to supeconducting have been detected well above their bulk $T_c$. If real, these signals could come from islands or regions of higher $T_c$ material stabilized by stacking faults, inclusions or other defects etc.. The abnormally high signals frequently disappear with time perhaps due to the slow decay of metastable regions or percolation paths. A typical example, one of many is the 250 K signal found in oxygen-annealed YBCO samples where non-linear-IV characteristics were observed that were sensitive to small magnetic fields [119]. At the time a determined attempt to track down the source of the signal was inconclusive. Today, when superconductors are more carefully prepared than they were in the early days of high $T_c$ when searching had a high priority, there are few if any reports of magnetic or resistive signals well above the bulk $T_c$. While this may be simply because the signals do not appear in the better samples, it may also be due to the fact that people are no longer looking.

●The third class is taken from reports of superconductivity in other than cuprate oxides. For instance Reich and Tsabba [ 120 ] have reported evidence for superconductivity at 90K (both electrical and diamagnetic signals) in $Na_xWO_3$. The signals are believed to emanate from the surface region of a single crystal that contains a metastable concentration gradient of Na achieved by annealing a single crystal of stable $NaWO_3$. The evaporation of Na during the annealing was estimated to leave a concentration gradient in the surface region of around 5% Na. Electron spin resonance experiments on the same material [121] was interpreted to give evidence for weak-coupled superconductivity in an unidentified minority phase. No evidence for 90K superconductivity has been reported from other laboratories suggesting that either the signals were spurious or that the superconducting phase requires very special conditions before it can be retained. It is worth exploring the metastable region of the phase diagram with more control using the advanced thin film techniques that are available. If substantiated the $T_c$ would be in a cubic perovskite structure with no magnetic ions and thus quite distinct from cuprate superconductivity.

## Figure Captions

FIG 1. The relationship between $T_c$ and the $CuO_2$-layer hole concentration, $p(CuO_2)$, in the $Bi_2Sr_2(Y_{1-x}Ca_x)Cu_2O_{8-d}$ system. $p(CuO_2)$ is taken as an average of the values determined for the $CuO_2$-plane hole concentration by coulometric redox analysis and by Cu $L3$-edge XANES spectroscopy. The actual cation doping level is ~2 times $p(CuO_2)$, because almost half of the holes migrate to the BiO layers. . The threshold hole concentration for the appearance of superconductivity is seen at $p(CuO_2)=0.06$ [from Karpp et al, 54].

FIG 2. A schematic phase diagram ($T_c$ versus superfluid density $p$ in the $CuO_2$ layer) that illustrates the enhancement of $T_c$ due to the insertion of charge reservoir layers that contain pairing centers. Thin dotted curve, the pairing amplitude (mean field); dotted curves; $T_0^{(1)}$ and $T_0^{(n)}$, the phase ordering temperature without and with the charge reservoir layers, respectively; Blue solid curve, $T_c$ of 214; Red dash-dot curve, $T_c$ of 2212. As a consequence of the suppression of fluctuations the model of Kivelson and Fradkin would predict that $p_{opt}$ should shift to the left (lower superfluid density). The results of Karppinen *et al.* are taken as evidence for this shift.

FIG 3. Structure of $Pr_2Ba_4Cu_7O_{15-d}$ (Pr247). Pr247 consists of the Pr123 unit ("1-2-3") and the Pr124 unit ("1-2-4"). In addition to two $CuO_2$ planes, the "1-2-3" ("1-2-4") contains a single chain (a double chain). The Cu atoms in the double chain do not form a "ladder" structure but a "zigzag" chain [from S. Sasaki et al, [72]].

FIG 4. (a) Temperature dependence of $1/T_1$ of the double-chain in Pr-247. Above $T_c$, the $T_1$ process exhibited a single-exponential time evolution, which yields a unique value of $T_1$. Below $T_c$, the $T_1$ process was reproduced by a bi-exponential function with two time constants, $T_1 S$ and $T_1 L$ indicating 20% of the chain copper nuclei belong to the superconducting phase. (b) Shows the antiferromagnetic magnetization of the two different copper oxide planes in the 247 unit cell [from S. Sasaki et al, [72]].

FIG 5. An ionic representation of the $CuO_2$ layer in the ab-plane. The squares that have no dashed lines represent the ground state of the undoped layer. Open circles represent ions with filled shells either O p6 and Cu d10; filled circles represent ions with open

shells, O p5 or Cu d9; a) charge transfer exciton; b) doped hole on oxygen; c) bound exciton-hole(eh) particle; d) extended bound exciton-hole particle; e) a higher energy (Vpp) configuration of the bound exciton-hole.

.

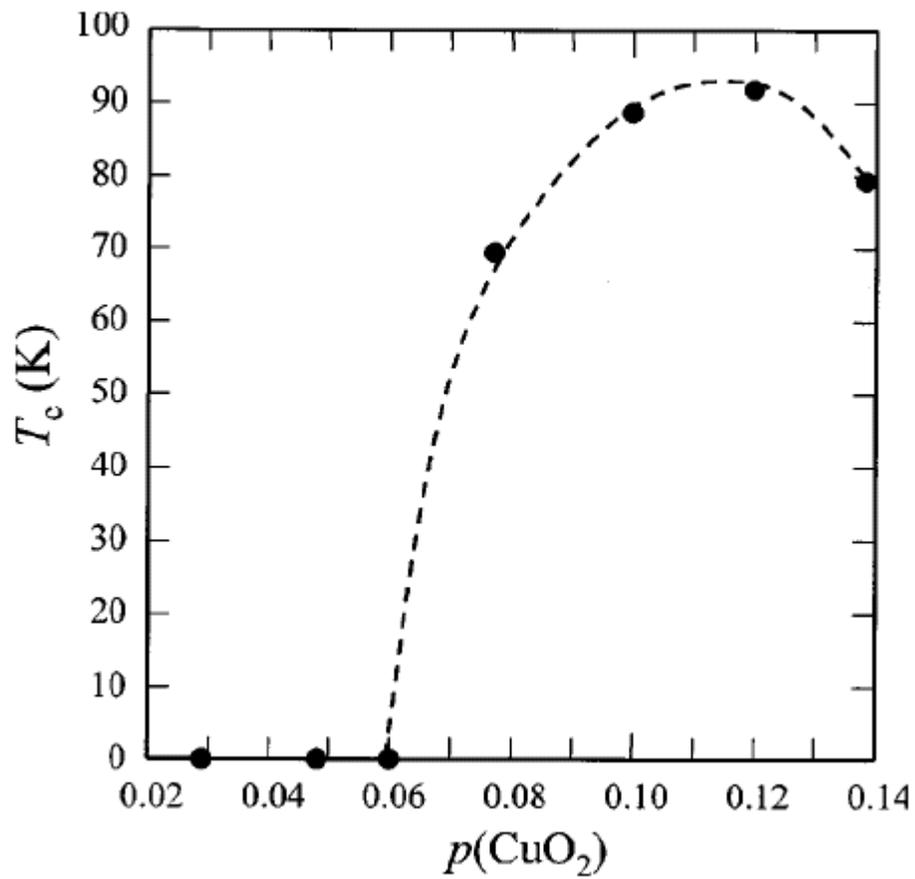

Fig. 1

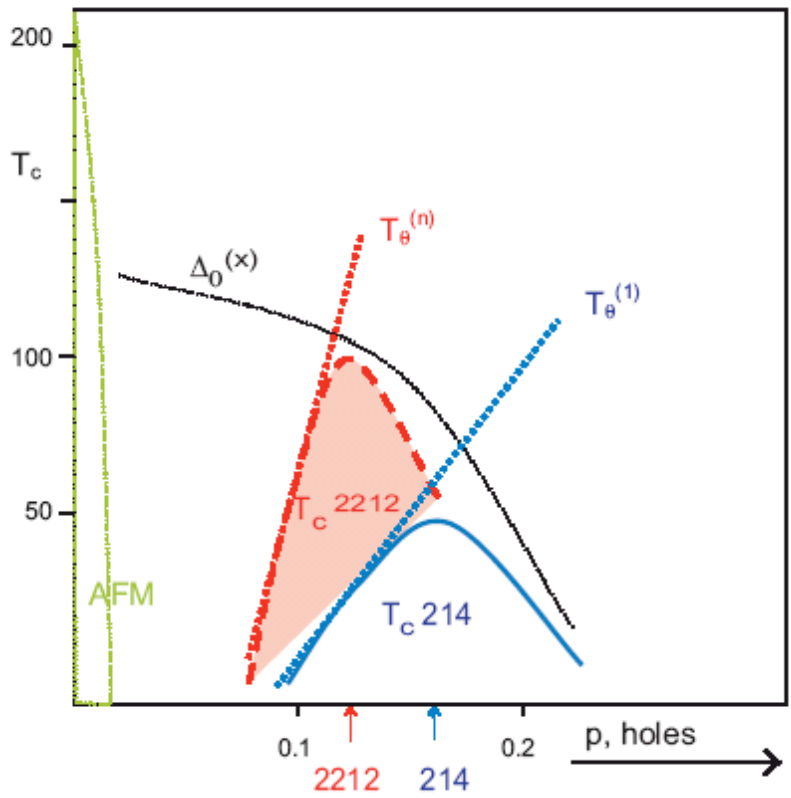

Fig. 2

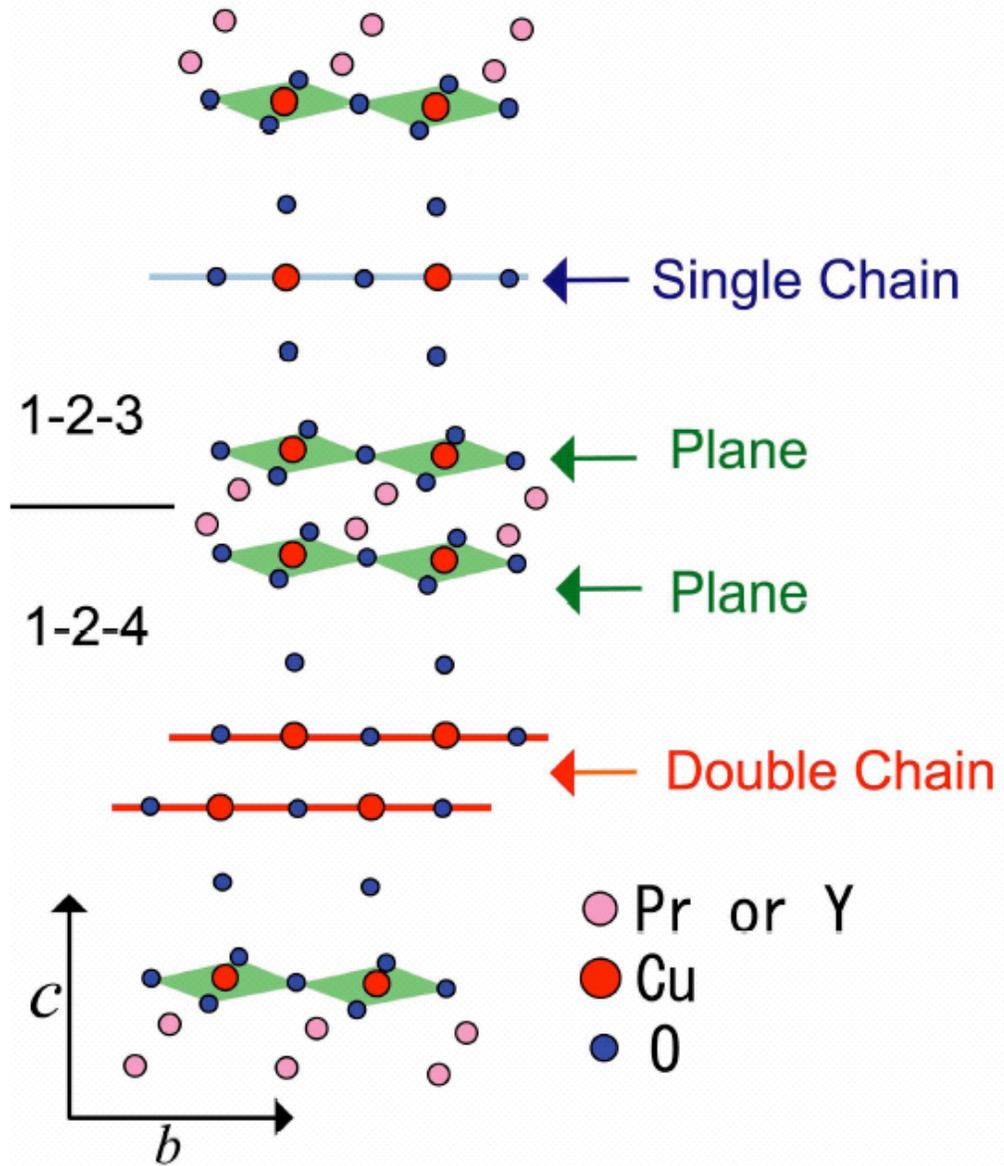

Fig. 3

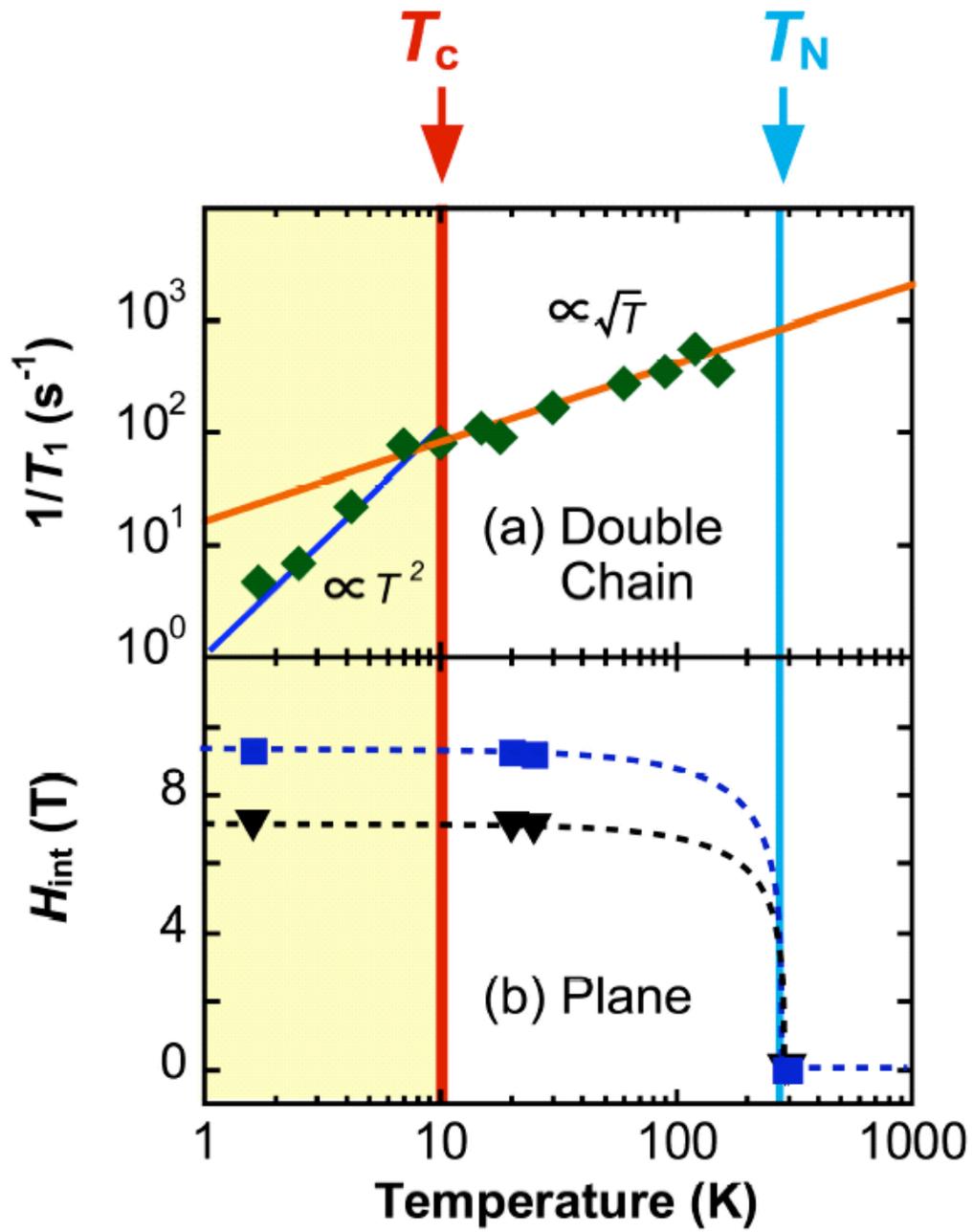

Fig. 4

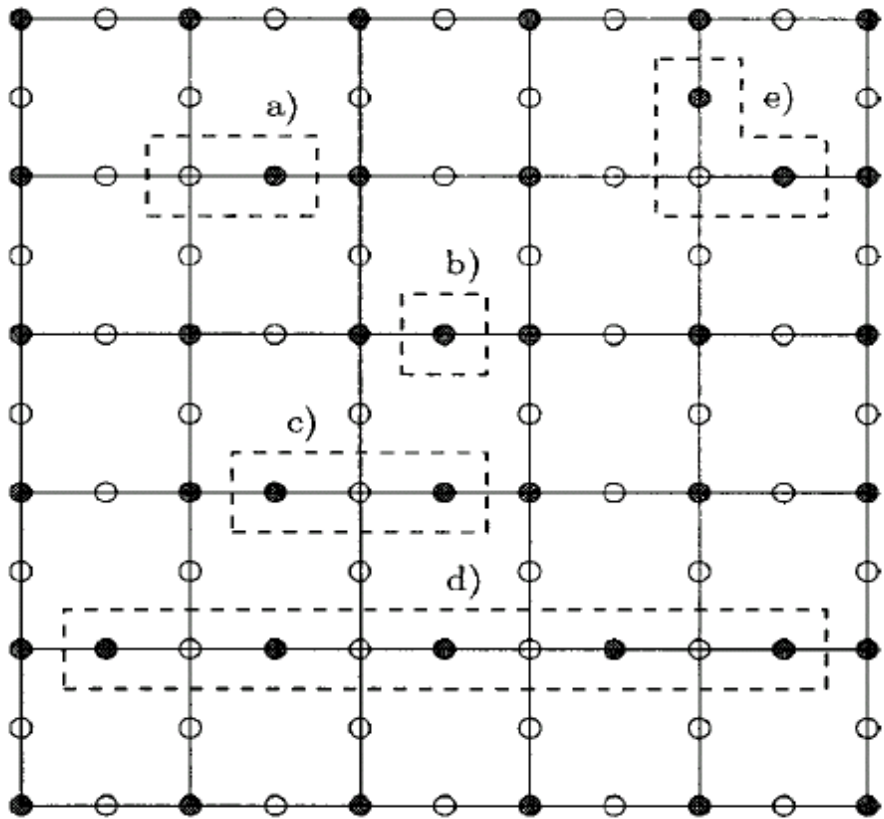

Fig. 5